\documentclass[print]{aa}
\usepackage{graphicx}
\usepackage{natbib}
\bibpunct{(}{)}{;}{a}{}{,}
\begin{document}

\title{Cooling of a rotating strange star in the color
superconducting phase with a crust}

\author{Y.W. Yu \and X.P. Zheng}

\institute{Institute of Astrophysics, Huazhong Normal University,
Wuhan 430079, China} \mail{ zhxp@phy.ccnu.edu.cn}
\date{Received 9 September 2005 / Accepted 14 December 2005}

\abstract
{}
{We investigate the thermal evolution of strange stars in the
2-flavor color superconductivity and color-flavor locked phases
under the influence of deconfinement heating.}
{Due to the spin-down of strange stars, the nuclear matter at the
base of the thin crusts dissolves into quarks, releasing energy to
heating the stars. On the other hand, the neutrino emissivities and
specific heat involving pairing quarks are suppressed by the large
pairing gap in color superconducting phases. Then the thermal
evolution equation of strange stars is calculated.}
{Deconfinement heating delays the cooling of strange stars
considerably. The presence of color superconductivity with a large
gap enhances this effect. Especially, in the color-flavor locked
phase, the stars cannot be very cold at an early age but they cool
slowly. For the stars with strong magnetic fields, a significant
heating period could exist during the first several ten or hundred
years. In addition, we reckoned that a possible theoretical limit
line, which is determined by the competition between deconfinement
heating and surface photon cooling, may indicate the upper limit
temperature that isolated compact stars should not exceed. }
{Deconfinement heating is important for the thermal evolution of
strange stars and is especially determinant for the stars in
color-flavor locked phase which could show characteristic cooling
behavior under this heating effect.}


\maketitle
\section*{1. Introduction}
Cooling simulation based on interior physics is of significant
interest for the research  of compact stars. In accordance with
nuclear physics, a quark matter core could be produced in the
interior of compact stars (hybrid star) and even strange quark
matter stars (strange star, SS) may exist. Phenomenological and
microscopic studies have confirmed that quark matter at a
sufficiently high density, as in compact stars, undergoes a phase
transition into a color superconducting state, which are typical
cases of the 2-flavor color superconductivity (2SC) and color-flavor
locked (CFL) phases \citep{Shovkovy(2004),Alford(2004)}. Theoretical
approaches also concur that the superconducting order parameter,
which determines the gap $\Delta$ in the quark spectrum, lies
between 10 and 100MeV for baryon densities existing in the interiors
of compact stars. Recently, the cooling of hybrid stars with color
superconducting quark cores have been investigated. Stars with CFL
cores behave similarly to ordinary neutron stars
\citep{Shovkovy(2002)}. And with a designed 2SC+X phase, hybrid
stars could also explain the cooling data properly \citep{Grigorian
et al(2005)}. In these cases, the thermal properties of the quark
cores are suppressed by the large gap, and the hadronic matter parts
of the stars play an important role in their cooling history.
\par
The thermal evolution of SSs has also been extensively discussed. In
early works, it was generally accepted that the surface temperature
of SSs should be lower than neutron stars at the same age due to the
quark direct Urca (QDU) processes
\citep{Alcock(1988),Pizzochero(1991),Page(1992),Schaab(1996)}.
However, since the electron fraction could be small or even vanish,
the QDU processes may be switched off. The cooling of SSs dominated
by the quark modified Urca (QMU) and quark bremsstrahlung (QB)
processes can be slower than neutron stars with standard cooling
\citep{Schaab(1997a),Schaab(1997b)}. Of course, a color
superconducting phase could occur in SSs, and its effect on the
cooling of the stars is a significant issue. \citet{Blaschke(2000)}
show that the cooling of the stars in the 2SC phase (2SS hereafter)
is compatible with existing X-ray data but that the stars in CFL
phase (CSS hereafter) cool down too rapidly, which disagrees with
the data. However, in those calculations an important factor, as
described below, is ignored.
\par
An SS, both in normal phase and in color superconducting phase, can
sustain a tiny nuclear crust with a maximum density below neutron
drip ($\sim10^{11}\rm g\hspace{0.1cm} cm^{-3}$) and mass
$M_{c}\leq10^{-5}M_{\odot}$ due to the existence of a strong
electric field on the quark surface
\citep{Alcock(1986),Usov(2004),Zheng(2006)}. The spin-down of the
star makes the matter at the bottom of the crust compress. As soon
as the density exceeds neutron drip, the surplus matter in the crust
falls into the quark core in the form of neutrons. Consequently, the
engulfed neutrons dissolve into quarks, and the released energy
during this process leads to a so-called deconfinement heating (DH).
\citet{Yuan(1999)} claim that DH delays the cooling of SSs in normal
phase (NSS) and may even lead to a slight increase in the
temperature at the early ages of a star under specific conditions.
\par
We argue that the heating effect on the thermal evolution of stars
in color superconducting phase is huge compared to NSSs because of
the suppression of specific heat and neutrino emission involving
pairing quarks. Therefore, focus on the effects of DH on the cooling
of 2SSs and CSSs in this paper. Our paper is arranged as follows. We
recall neutrino emissivities and specific heat, color
superconductivity, and the DH mechanism in Scets.2, 3, and 4,
respectively. The cooling curves and the corresponding explanations
are presented in Sect.5. Section 6 contains our conclusion and
discussions.
\section*{2. Neutrino emissivities and specific heat}
The emissivity associated with the QDU processes $
d{\rightarrow}ue\bar{\nu}$ and $ue{\rightarrow}d{\nu}$ of quarks is
\citep{Iwamoto(1982)}
\begin{equation}
{\epsilon}^{(D)}{\simeq}8.8{\times}10^{26}
{\alpha}_c\left({\frac{{\rho}_b}{{\rho}_0}}\right)Y_e^{1/3}T_9^6
{\rm \hspace{0.1cm} erg \hspace{0.1cm} cm^{-3} \hspace{0.1cm}
sec^{-1}},
\end{equation}
where ${\alpha}_c$ is the strong coupling constant, ${\rho}_b$ is
baryon density and ${\rho}_0=0.17{\rm \hspace{0.1cm} fm^{-3}}$ the
nuclear saturation density, and $T_9$ is the temperature in units of
$10^9\hspace{0.1cm} {\rm K}$. The electron fraction
$Y_e={\rho}_e/{\rho}_b$ is small and even vanishes at a certain set
of parameters of ${\rho}_b$, ${\alpha}_c$ and s-quark mass $m_s$
\citep{Duncan(1983)}. It has not yet been considered that the
contribution to the emissivity by the \textit{s-u} reaction, which
is suppressed by an extra factor $\rm sin^{2}\theta_{c}\sim10^{-3}$
compared to the \textit{d-u} reaction \citep{Duncan(1983)}, where
$\theta_{c}$ is the Cabibbo angle. When the QDU processes are
switched off due to a small electron fraction
($Y_{e}<Y_{ec}=(3/\pi)^{1/2}m_{e}^{3}\alpha_{c}^{-3/2}/64$. ), the
contribution to the emissivities by the QMU processes
$dq{\rightarrow}uqe\bar{\nu}$ and QB processes
$q_1q_2{\rightarrow}q_1q_2{\nu}\bar{\nu}$ dominate. These
emissivities were also estimated as \citep{Iwamoto(1982)}
\begin{equation}
{\epsilon}^{(M)}{\simeq}2.83{\times}10^{19}
{\alpha}_c^2\left({\frac{{\rho}_b}{{\rho}_0}}\right)T_9^8{\rm
\hspace{0.1cm} erg \hspace{0.1cm} cm^{-3} \hspace{0.1cm}
sec^{-1}},
\end{equation}
\begin{equation}
{\epsilon}^{(B)}{\simeq}2.98{\times}10^{19}
\left({\frac{{\rho}_b}{{\rho}_0}}\right)T_9^8{\rm \hspace{0.1cm}
erg \hspace{0.1cm} cm^{-3} \hspace{0.1cm} sec^{-1}}.
\end{equation}
In order to compute the thermal evolution of the stars, we also need
the the specific heat involving quarks and electrons written as
\citep{Iwamoto(1982),Blaschke(2000)}
\begin{equation}
c_q{\simeq}2.5{\times}10^{20}\left({\frac{{\rho}_b}{{\rho}_0}}\right)^{2/3}T_9{\rm
\hspace{0.1cm} erg \hspace{0.1cm} cm^{-3}\hspace{0.1cm} K^{-1}},
\end{equation}
\begin{equation}
c_e{\simeq}0.6{\times}10^{20}\left({\frac{{Y_e{\rho}_b}}{{\rho}_0}}\right)^{2/3}T_9{\rm
\hspace{0.1cm} erg \hspace{0.1cm} cm^{-3}\hspace{0.1cm} K^{-1}}.
\end{equation}
\par
Since the mass of the crust is very small, $M_{\rm
c}{\leq}10^{-5}M_{\odot}$, compared with the total mass of the star,
its contribution to neutrino emissivity and specific heat can be
neglected \citep{Gudmundsson et al(1983),Lattimer(1994)}. Here we
also ignore the neutrino emissivity and specific heat due to the
photon-gluon excitation, because this excitation is only important
for a temperature higher than $70{\rm MeV}$ \citep{Blaschke(2000)},
which is much higher than the typical temperature in our
calculation.
\section*{3. Color superconductivity}
\par
It is widely accepted that the color superconducting phase is the
real ground state of quantized chromodynamics at asymptotically
large densities. At a certain range of the quark chemical potential
the quark-quark interaction is attractive, driving the pairing
between the quarks
\citep{Alford(1998),Alford(1999),Alford(2003),Rapp(1998),Shovkovy(2004)}.
Because of the pairing, QDU processes are suppressed by a factor
${\rm exp}(-{\Delta}/k_{B}T)$, and QMU \& QB processes are
suppressed by a factor ${\rm exp}(-2{\Delta}/k_{B}T)$ for
$T<T_c{\simeq}0.4\Delta/k_{B}$ \citep{Blaschke(2000)}. In the 2SC
phase, two color states of \textit{u} and \textit{d} quarks pair,
whereas the \textit{s} quark is unpaired. To be specific, we suppose
that blue-green and green-blue \textit{u-d} quarks are paired ,
whereas red \textit{u} and \textit{d} quarks ($u_{r},d_{r}$) remain
unpaired. As a consequence, the QDU processes on the red (unpaired)
quarks, as $d_{r}\rightarrow u_{r}e \bar{\nu}$, as well as QMU,
$d_{r}q_{r}\rightarrow u_{r}q_{r}e \bar{\nu}$, and QB,
$q_{1r}q_{2r}\rightarrow q_{1r}q_{2r}\bar{\nu}\nu $, are not
blocked, whereas other processes involving paired quarks are blocked
out by a large pairing gap. Therefore, it can be estimated that the
neutrino emissivities in 2SC phase are reduced by about one
magnitude \citep{Blaschke(2000)}. On the other hand, the neutrino
processes involving all flavors are suppressed in the CFL phase by
the exponential factors. For both phases where the specific heat
contributed by the paired quarks is also changed, we apply the
formula \citep{Blaschke(2000)}
\begin{equation}
\begin{array}{cc}
c_{sq}=3.2c_q\left({\frac{T_c}{T}}\right)\times\hspace{5cm}\\
\left[2.5-1.7\left({\frac{T}{T_c}}\right)+3.6\left({\frac{T}{T_c}}\right)^2\right]{\rm
exp}\left(-\frac{\Delta}{k_{B}T}\right).
\end{array}
\end{equation}

\section*{4. Deconfinement heating}
The effect of DH is determined by the number of neutrons engulfed by
the quark core, in other words, the variation in the mass of the
crust. The total heat released per time unit as a function of $t$ is
\begin{equation}
H_{\rm dec}(t)=-q_n\frac{1}{m_{b}}\frac{d M_{\rm c}}{d
{\nu}}\dot{\nu},
\end{equation}
where $q_n$, the heat release per absorbed neutron, is expected to
be in the range $q_n{\sim}10-30{\rm MeV}$ \citep{Haensel(1991)}, and
$m_{b}$ is the mass of baryon. Assuming the spin-down is induced by
the magnetic dipole radiation, the evolution of the rotation
frequency $\nu$ is given by
\begin{equation}
\dot{\nu}=-\frac{8\pi^{2}}{3Ic^3}{\mu}^2{\nu}^3{\rm sin}^2{\theta},
\end{equation}
where $I$ is the stellar moment of inertia, ${\mu}=\frac{1}{2}BR^3$
is the magnetic dipole moment, and $\theta$ is the inclination angle
between magnetic and rotational axes. The mass of the crust $M_{\rm
c}$ is calculated by a quadratic function of $\nu$ by
\citet{Glendenning(1992)}, whose result describes the cases with
intermediate frequencies ($\leq500\rm Hz$) very well.
\citet{Zdunik2001} improve the calculation nearly up to the
Keplerian frequency using a polynomial including terms of higher
order in $\nu$. The mass of the crust reads \citep{Zdunik2001}
\begin{equation}
M_{\rm c}=M^{0}_{\rm c}(1+0.24\nu^{2}_{3}+0.16\nu^{8}_{3}),
\end{equation}
where $\nu_{3}=\nu/10^{3}$Hz, and $M^{0}_{\rm
c}\approx10^{-5}M_{\odot}$ is the mass of the crust in the static
case.
\section*{5. Cooling curves}
The thermal evolution with DH of a star is determined by the
equation
\begin{equation}
C\frac{d T}{d t}=-L_{\nu}-L_{\gamma}+H_{\rm dec},
\end{equation}
where $C$ is the total specific heat, $L_{\nu}$ the neutrino
luminosity, and $L_{\gamma}$ the surface photon luminosity given by
\begin{equation}
L_{\gamma}=4{\pi}R^2{\sigma}T_s^4,
\end{equation}
where ${\sigma}$ is the Stefan-Boltzmann constant and $T_s$ the
surface temperature. The internal structure of SSs can be regarded
as temperature-independent \citep{Glen(1980)}, and the surface
temperature is related to internal temperature by a coefficient
determined by the scattering processes occurring in the crust. In
the work of \citet{Blaschke(2000)}, this relation is given by a
simple expression $T_{s}=5\times10^{-2}T$, but following
\citet{Yuan(1999)}, we apply an accurate formula that is
demonstrated by \citet{Gudmundsson et al(1983)},
\begin{equation}
T_{s}=3.08\times10^6g_{s,14}^{1/4}T_{9}^{0.5495},
\end{equation}
where $g_{s,14}$ is the proper surface gravity of the star in units
of $10^{14}\rm cm\hspace{0.1cm}s^{-2}$; or for a recent version see
the result of \citet{Potekhin(1997)}.
\par
In our calculations, to be specific, we consider a model of
canonical SS of $1.4M_{\odot}$ at a constant density, which is a
very good approximation for SSs of mass $M\leq1.4M_{\odot}$
\citep{Alcock(1986)}. As used by \cite{Blaschke(2000)}, we take
$Y_{e}=10^{-5}$, $\alpha_{c}=0.25$, $\rho=3\rho_{0}$ for
$Y_{e}>Y_{ec}$, which is a representative set of parameters for
which the QDU processes contribute to the cooling, whereas
$Y_{e}=0$, $\alpha_{c}=0.15$, $\rho=5\rho_{0}$ for $Y_{e}<Y_{ec}$.
And we also choose $\Delta=100$MeV and $q_n=20{\rm MeV}$, the
initial temperature $T_0=10^9{\rm K}$, initial period $P_0=0.78{\rm
ms}$, and the magnetic tilt angle $\theta=45^{\circ}$. The
gravitational red-shift is also taken into account. Then the
effective surface temperature detected by a distant observer is
$T_{s}^{\infty}=T_{s}\sqrt{1-R_{g}/R}$, where $R_{g}$ is the
gravitational stellar radius.
\par
We plot the cooling curves without DH of NSSs (solid curves), 2SSs
(dotted curves), and CSSs (dashed curves) in Fig.1. And the
observational data, which are taken from Tables 1 and 2 in
\citet{Page et al(2004)}, are also shown in order to give the
readers a feeling of the position of the illustrative curves in the
logarithm $T_s^{\infty}-t$ plane. But we will not try to fit the
data carefully in this paper. It shows that the cooling history of
2SSs are similar to NSSs, whereas CSSs become very cold at an early
age since the specific heat is very small, cooler than $10^{4.5}$K
after 1000yr for $Y_{e}=10^{-5}$ or several hours for $Y_{e}=0$
(this curve is not shown in Fig.1). We can see the curves of CSSs
are very far from the data. These conclusions are also indicated by
\citet{Blaschke(2000)} using the relation of
$T_{s}=5\times10^{-2}T$.
\par
\begin{figure}
\resizebox{\hsize}{!}{\includegraphics{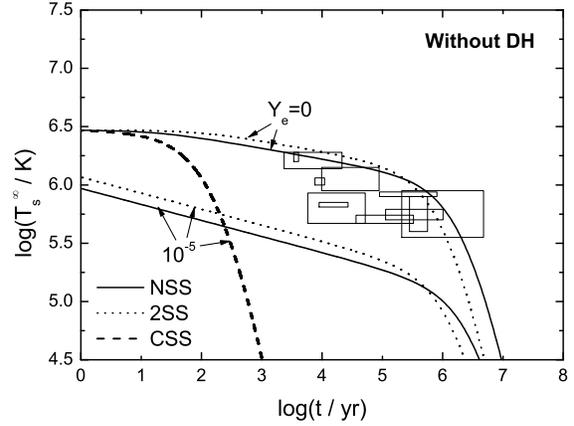}}
\caption{Cooling curves of NSSs, 2SSs, and CSSs without DH.}
\end{figure}
%
%
%
\begin{figure}
\resizebox{\hsize}{!}{\includegraphics{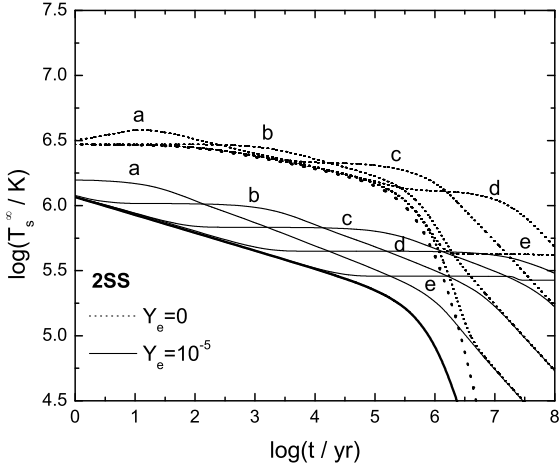}}
\caption{Cooling curves of 2SSs with DH for various magnetic fields
(curve a: $10^{12}$G, b: $10^{11}$G, c: $10^{10}$G, d: $10^{9}$G, e:
$10^{8}$G) and the curves without DH (thick curve).}
\end{figure}
%
Figure 2 shows the cooling behaviors of 2SSs with DH for various
magnetic fields ($10^{8}-10^{12}$G). And the analogs for NSSs can be
seen in \citet{Yuan(1999)}. We can see DH delays the stellar cooling
considerably. As discussed by \citet{Yuan(1999)}, the stronger the
magnetic field the more rapid the spin-down, and most of the nuclear
matter in the crust dissolves during an earlier and shorter time.
For a 2SS with a strong ($B=10^{12}$G) field and small electron
fraction (dotted curve a), a distinct heating period exists in the
first several ten years. And in the cases of weak fields
($B<10^{10}$G), stars could maintain high temperatures even at older
ages ($>10^{6}$yrs). In the following paragraph, we discuss in
detail how DH induces temperature rise and delays cooling.
\par
\begin{figure}
\resizebox{\hsize}{!}{\includegraphics{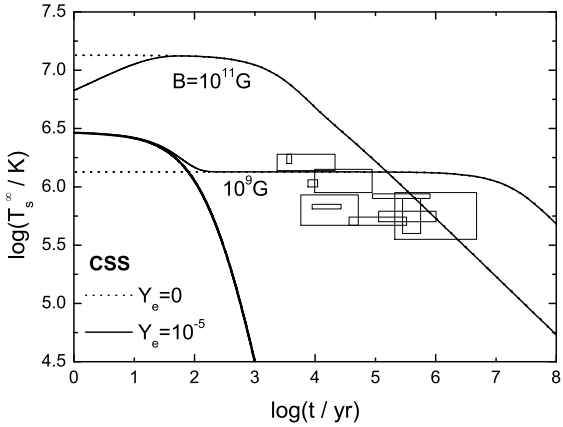}}
\caption{Cooling curves of CSSs with DH for a strong ($B=10^{11}$G)
and a weak ($B=10^{9}$G) magnetic field and the curve without DH
(thick curve).}
\end{figure}
%
We here pay more attention to the situation of CSSs because we argue
the existence of a marked heating effect relative to the reduced
emissions. To be clear, Fig. 3 shows the cooling curves of CSSs with
DH for both a strong ($B=10^{11}$G) and a weak ($B=10^{9}$G)
magnetic field. It is obvious that the cooling curves are changed
dramatically by DH. The strong magnetic field induces a rapid
spin-down of the star at the earliest ages, which could enhance the
effect of DH to make it greater than the cooling effect at the
beginning. As a result, the temperature should rise due to the
surplus heat until the increasing luminosity equals the heating
effect, $L_{\gamma}=H_{\rm dec}$, so a net heating period appears at
the earliest ages. On the other hand, the temperature of the star
with a weak field decreases but does not rise, due to the relative
greater cooling effect at the start until the thermal release is
compensated for by DH entirely: $L_{\gamma}=H_{\rm dec}$. In this
case, since the confinement energy deposited in the crust is
released slowly to heat the star, the star with a weak field can
maintain a high temperature even at older ages ($> 10^{6}{\rm
yrs}$). To conclude, both the stars with strong and weak fields,
after several hundred years (the specific value of the time is
determined by the specific condition of the star), could arrive at
an equilibrium between the cooling and heating effects. From then
on, the temperature could only be reduced in order to rebuild the
equilibrium when $H_{\rm dec}$ deceases with time, so the cooling of
the stars is delayed. Due to this delay, the curves of CSSs cannot
be in conflict with observational data as shown in Fig.3. Since the
neutrino emission involving all quarks in CSSs is suppressed, the
equilibrium discussed above is only determined by DH and photon
emission, and has hardly anything to do with the interior thermal
properties of the star. Therefore, the dependence of the cooling on
the electron fraction is eliminated after the first several hundred
years. Going back to Fig.2, we see that the mechanism described
above also influences the cooling of 2SSs. However, since the
equilibrium ($L_{\nu}+L_{\gamma}=H_{\rm dec}$) should involve
neutrino luminosity, which is larger at high temperature whereas
smaller at low temperature than photon luminosity, the evolution of
2SSs may be more complicated than it is for CSSs (see Fig.4 for
detail), i.e., the cooling history can be roughly divided into a
neutrino cooling stage ($L_{\nu}\gg L_{\gamma}$) and a photon
cooling stage ($L_{\nu}\ll L_{\gamma}$), and the cooling is
sensitive to the electron fraction; and the moment when the
equilibrium is achieved could be very different for different
magnetic fields.
\par
\begin{figure}
\resizebox{\hsize}{!}{\includegraphics{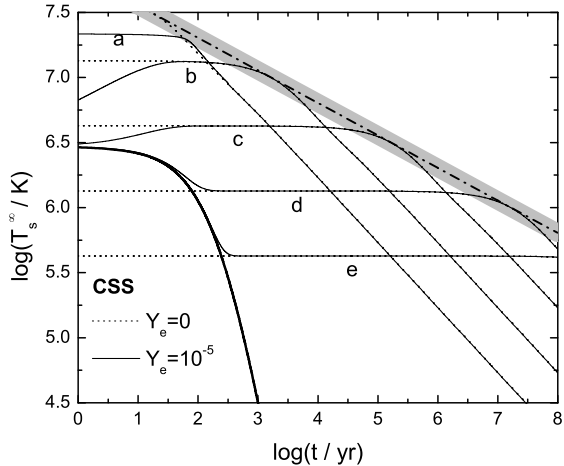}} \caption{The
same as Fig.2 but for CSSs. The dash-dotted curve is the limit line,
and the band-like structure is supposed to reflect the uncertainty
of $q_n$.}
\end{figure}
%
Figure 4 shows the cooling curves of CSSs with different magnetic
fields. We can see the cooling of the stars with any field is
delayed, just as we find in Fig.2. However, in comparison with
Fig.2, there is a question of why the temperature rise of CSSs can
be more significant than the one of 2SSs. For both CSSs and 2SSs, as
discussed in the previous paragraph, the reason for the temperature
rise is that the heating effect is greater than the cooling effect
at the beginning: $H_{\rm dec}>L_{\gamma}$ for CSSs and $H_{\rm
dec}>L_{\nu}+L_{\gamma}$ for 2SSs. Since the neutrino term is absent
for CSSs, the initial difference between the heating term and
luminosity of CSSs is much larger than the one of 2SSs with the same
magnetic field and initial temperature. On the other hand, with the
rise in temperature, the increase in the luminosity is proportional
to $T^{2.2}$ for CSSs (see Eqs.(11, 12)) but to $T^{8}$ for 2SSs
with a small electron fraction (see Eqs.(2, 3), where the term of
photon luminosity is ignored since $L_{\nu}\gg L_{\gamma}$ at high
temperature). Therefore, 2SSs can achieve the equilibrium easily
after a comparatively small temperature rise, but the magnitude of
the rise for CSSs needs to be much larger. In addition, the needed
magnetic field intensity to induce the temperature rise of CSSs
could be smaller than 2SSs since the initial cooling effect of CSSs
is smaller.
\par
We link the points where the cooling curves turn down together as a
line (dash-dotted curve) in logarithm $T_{s}^{\infty}-t$ plane in
Fig.4. The temperature indicated by the line is expressed as a power
form $T^{\infty}_{s,\rm lim}=6.4{\times}10^{7}t^{-1/4}{\rm K}$. It
may be well-founded that any other cooling curves of isolated stars
(to our knowledge), regardless of star models, will be below this
line due to the high heating effect and low cooling effect in CSSs.
Hence we reckon that the indicated temperature may be the upper
limit to what compact stars can reach at a given age. We must
emphasize that this line found from the cooling curves of CSSs is
only determined by the equilibrium between DH and surface photon
cooling.
\begin{figure}
\resizebox{\hsize}{!}{\includegraphics{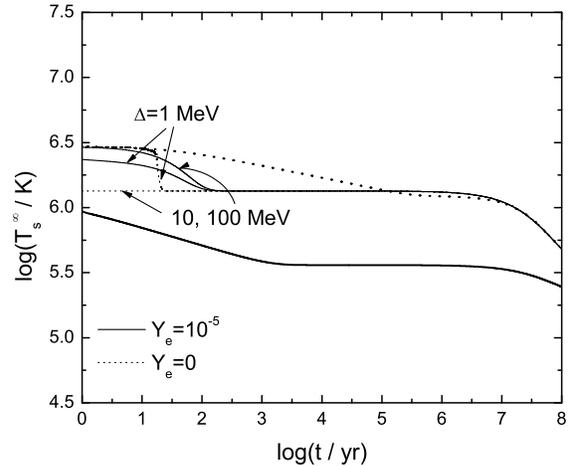}}
\caption{Cooling curves of CSSs with $B=10^9{\rm G}$ for different
gaps. The thick curves correspond to NSSs.}\
\end{figure}
%
\par
Finally, we present the cooling curves of CSSs for different gaps in
Fig.5. It can be seen that the cooling curves are almost independent
of the gap on a very large parameter scale (${\Delta{\sim}10-100{\rm
MeV}}$).
\section*{6. Conclusion and discussions}
We have studied the cooling behaviors of rotating SSs in the
presence of color superconductivity by considering the effect of DH.
The thermal evolution of SSs is now quite different from previous
results, because DH can delay the cooling, and color
superconductivity enhances this effect significantly, especially in
the CFL phase. For CSSs, the previous discussions point out that the
specific heat is determined by electrons since the contribution of
quarks has been suppressed. This reduction leads to a very rapid
cooling that disagrees with observational data
\citep{Blaschke(2000)}. However, when we consider the effect of DH,
the results should imply that the cooling curves could not be in
serious conflict with the data. We even find it is possible that
CSSs reach a higher temperature than other kinds of compact stars in
their cooling history. The limit temperature line should illustrate
this conjecture.
\par
To be specific, as pointed out by \citet{Yuan(1999)}, a temperature
rising period could exist at the early ages due to the DH with a
strong magnetic field. And we argue that the presence of color
superconductivity may lead to a significant rise. \citet{Yuan(1999)}
propose that this phenomena may be a signature of the existence of
SS. In our opinion, if the theory of the color superconductivity is
reliable, observing a young and quite hot source may be possible,
although we also note that an important so-called brightness
constraint has been suggested recently by \citet{Grigorian(2005)},
who argues that it is unlikely that objects with a given age are
hotter than those already observed. Statistically, this constraint
is a good finding, but we think that in theory the possibility of
the existence of young hotter stars still cannot be rejected
absolutely. Of course, for our model, the early evolution also may
be changed to a certain extent if we consider the formation of the
crust of the star. On the other hand, for those stars with weak
fields ($<10^{10}$G), our results show that they can maintain a high
temperature at older ages ($>10^{6}$yrs). Unfortunately, these older
sources with weak fields also have been not detected up to now (see,
for example, \citet{Popov et al(2003)} for the list of close-by
cooling pulsars). To summarize, at the present point of
observations, there is no evidence of the existence of extra hot
sources. It may imply that the model needs some further
improvements. However, it still should be emphasized that the
various heating mechanisms in compact stars need to be given more
importance when we talk about the star's cooling when using the
so-called standard scenario.

\begin{acknowledgements} We would like to thank Prof. D. F. Hou
for the useful discussion. We are especially indebted to the
anonymous referee for his/her useful comments that helped us to
improve the paper. This work was supported by the NFSC under Grant
Nos. 10373007 and 90303007.
\end{acknowledgements}

\end{document}